\documentclass[10pt,prl,onecolumn,preprint,floats,floatfix,aps,nofootinbib,preprintnumbers]{revtex4-1}
\usepackage{graphicx}
\usepackage{amssymb}
\usepackage{dcolumn}
\usepackage{amsmath}
\usepackage{slashed}
\usepackage{multirow}
\usepackage{natbib}
\usepackage{color}
\usepackage{ulem}

\usepackage{tikz}
\usetikzlibrary{arrows,decorations.pathmorphing}

\newcommand{\sD}[1]{\begin{tikzpicture}[scale={#1},photon/.style={decorate,decoration={snake,post length=1mm}}]
        \draw (0,4) -- (0.5,2);
        \draw[photon] (0,4) -- (0.5,2);
        \draw node[anchor=east] at (0,4) {\Large $n$};
        \draw (0,0) -- (0.5,2);
        \draw[photon] (0,0) -- (0.5,2);
        \draw node[anchor=east] at (0,0) {\Large $n$};
        \draw (0.5,2) -- (3.5,2);
        \draw (3.5,2) -- (4,4);
        \draw[photon] (3.5,2) -- (4,4);
        \draw node[anchor=west] at (4,4) {\Large $n$};
        \draw (3.5,2) -- (4,0);
        \draw[photon] (3.5,2) -- (4,0);
        \draw node[anchor=west] at (4,0) {\Large $n$};
      \node[circle,fill=darkgray,draw=black,inner sep=0pt,minimum size=0.1cm] at (0.5,2) {};
      \node[circle,fill=darkgray,draw=black,inner sep=0pt,minimum size=0.1cm] at (3.5,2) {};
      \draw node[anchor=south] at (2,2) {\Large $x$};
\end{tikzpicture}}

\newcommand{\tD}[1]{\begin{tikzpicture}[scale={#1},photon/.style={decorate,decoration={snake,post length=1mm}}]
        \draw (0,4) -- (2,3.5);
        \draw[photon] (0,4) -- (2,3.5);
        \draw node[anchor=east] at (0,4) {\Large $n$};
        \draw (0,0) -- (2,0.5);
        \draw[photon] (0,0) -- (2,0.5);
        \draw node[anchor=east] at (0,0) {\Large $n$};
        \draw (2,0.5) -- (2,3.5);
        \draw (2,3.5) -- (4,4);
        \draw[photon] (2,3.5) -- (4,4);
        \draw node[anchor=west] at (4,4) {\Large $n$};
        \draw (2,0.5) -- (4,0);
        \draw[photon] (2,0.5) -- (4,0);
        \draw node[anchor=west] at (4,0) {\Large $n$};
      \node[circle,fill=darkgray,draw=black,inner sep=0pt,minimum size=0.1cm] at (2,3.5) {};
      \node[circle,fill=darkgray,draw=black,inner sep=0pt,minimum size=0.1cm] at (2,0.5) {};
      \draw node[anchor=east] at (2,2) {\Large $x$};
\end{tikzpicture}}

\newcommand{\uD}[1]{\begin{tikzpicture}[scale={#1},photon/.style={decorate,decoration={snake,post length=1mm}}]
        \draw (0,4) -- (2,3.5);
        \draw[photon] (0,4) -- (2,3.5);
        \draw node[anchor=east] at (0,4) {\Large $n$};
        \draw (0,0) -- (2,0.5);
        \draw[photon] (0,0) -- (2,0.5);
        \draw node[anchor=east] at (0,0) {\Large $n$};
        \draw (2,0.5) -- (2,3.5);
        \draw (2,3.5) -- (4,0);
        \draw[photon] (2,3.5) -- (4,0);
        \draw node[anchor=west] at (4,0) {\Large $n$};
        \draw (2,0.5) -- (4,4);
        \draw[photon] (2,0.5) -- (4,4);
        \draw node[anchor=west] at (4,4) {\Large $n$};
      \node[circle,fill=darkgray,draw=black,inner sep=0pt,minimum size=0.1cm] at (2,3.5) {};
      \node[circle,fill=darkgray,draw=black,inner sep=0pt,minimum size=0.1cm] at (2,0.5) {};
      \draw node[anchor=east] at (2,2) {\Large $x$};
\end{tikzpicture}}

\newcommand{\sgD}[1]{\begin{tikzpicture}[scale={#1},photon/.style={decorate,decoration={snake,post length=1mm}}]
        \draw (0,4) -- (2,2);
        \draw[photon] (0,4) -- (2,2);
        \draw node[anchor=east] at (0,4) {\Large $n$};
        \draw (0,0) -- (2,2);
        \draw[photon] (0,0) -- (2,2);
        \draw node[anchor=east] at (0,0) {\Large $n$};
        \draw node[anchor=west] at (4,4) {\Large $n$};
        \draw (2,2) -- (4,0);
        \draw[photon] (2,2) -- (4,0);
        \draw node[anchor=west] at (4,0) {\Large $n$};
        \draw (2,2) -- (4,4);
        \draw[photon] (2,2) -- (4,4);
      \node[circle,fill=darkgray,draw=black,inner sep=0pt,minimum size=0.1cm] at (2,2) {};
      \node[circle,fill=darkgray,draw=black,inner sep=0pt,minimum size=0.1cm] at (2,2) {};
\end{tikzpicture}}

\newcommand{\gae}{\begin{array}{c}\,\sim\vspace{-28pt}\\> \end{array}}

\begin{document}

\title{Scattering Amplitudes of Massive Spin-2 Kaluza-Klein States \\Grow Only as ${\cal O}(s)$}
\author{R. Sekhar Chivukula$^{a,b}$}
\author{Dennis Foren$^{a,b}$}
\author{Kirtimaan A Mohan$^{b}$}
\author{Dipan Sengupta$^{a}$}
\author{Elizabeth H. Simmons$^{a,b}$}
\affiliation{$^{a}$ Department of Physics and Astronomy, 9500 Gilman Drive,
 University of California, San Diego }
 \affiliation{$^{b}$ Department of Physics and Astronomy, 567 Wilson Road, Michigan State University, East Lansing}


\begin{abstract}
{We present the results of the first complete calculation of the tree-level $2\to 2$ high-energy scattering amplitudes of the longitudinal modes of massive spin-2 Kaluza-Klein states, both in the case where the internal space is a torus and in the Randall-Sundrum model where the internal space has constant negative curvature. While individual contributions to this amplitude grow as ${\cal O}(s^5$), we demonstrate explicitly that intricate cancellations occur between different contributions, reducing the growth  to ${\cal O}(s)$, a slower rate of growth than previously argued in the literature. These cancellations require subtle relationships between the masses of the Kaluza-Klein states and their interactions, and reflect the underlying higher-dimensional diffeomorphism invariance. Our results provide fresh perspective on the range of validity of (effective) field theories involving massive spin-2 KK particles, with significant implications for the theory and phenomenology of these states.}
\end{abstract}
\maketitle

\section{Introduction}
In this letter we present the results of the first complete calculation of the tree-level $2 \to 2$ high-energy scattering amplitudes of the longitudinal polarizations of massive spin-2 Kaluza-Klein (KK) states in compactified five-dimensional theories. Fundamental or effective field theories (EFT) with massive spin-2 particles can arise in a variety of contexts, including alternative theories of gravity, string theory, and the AdS/CFT correspondence \cite{Maldacena:1997re,Witten:1998qj,Aharony:1999ti}, or through the compactification of Einstein gravity in higher dimensions (see \cite{Hinterbichler:2011tt,deRham:2014zqa} and references therein).  Massive spin-2 particles are also the object of LHC searches and are incorporated into phenomenologically-motivated models of particle physics and dark matter (for example, see \cite{Chivukula:2017fth} and references therein). In all of these cases the energy range in which calculations involving massive spin-2 particles are valid is determined by the rate of growth of the scattering amplitudes among the longitudinal polarization of these states; the faster the growth, the lower the energy scale at which unitarity is violated and  the (effective) theory becomes invalid. Prior work in the literature had argued that the rate of growth should be at least ${\cal O}(s^3)$; our explicit calculation proves that the rate is, instead, merely ${\cal O}(s)$, pushing the scale of unitarity violation higher. Our results therefore have significant implications for the theory and phenomonology of massive spin-2 states.

Constructing consistent theories of massive spin-2 particles presents several challenges. First, even without interactions, there are two linearly independent Lorentz-invariant mass terms which can be used, and only the specific combination introduced by Fierz and Pauli \cite{Fierz:1939ix} avoids propagating ghost degrees of freedom in flat spacetime \cite{vanDam:1970vg}. Second, the helicity-1 and (longitudinal) helicity-0 states of a massive spin-2 particle correspond to polarization tensors that are, at energies large compared to the mass of the particle, proportional to positive powers of that particle's momentum. These helicities cause contributions to the scattering amplitudes of massive spin-2 particles to grow rapidly with $s$, the squared center-of-mass energy. 

In particular, when the interactions of a spin-2 state are determined by a weak-field approximation of the four-dimensional (4D) Einstein-Hilbert (EH) action \cite{DeWitt:1967uc,Berends:1974gk}, naive power-counting suggests the elastic scattering amplitude for longitudinal massive spin-2 modes will grow like ${\cal O}(s^7)$.
Diffeomorphism invariance of the EH action, however, softens this high-energy behavior to ${\cal O}(s^5)$ --- a feature manifest in ``theory space" \cite{ArkaniHamed:2002sp}, where the helicity-1 and -0 states emerge as Goldstone bosons of the broken diffeomorphism invariance and in which power-counting is simple. As is customary, we define a scale $\Lambda_\lambda = (m_g^{\lambda-1} M_{Pl})^{1/\lambda}$ where $m_g$ is the mass of the scattered spin-2 particle and $M_{Pl}$ is the Planck scale associated with the 4D EH interactions. The scale $\Lambda_\lambda$ typically accompanies $s^\lambda$-like growth of a massive spin-2 scattering amplitude, and so an effective theory with a single massive spin-2 particle will typically have a cutoff scale of order $\Lambda_5$. There exist deformations of the theory \cite{ArkaniHamed:2002sp,ArkaniHamed:2003vb,Schwartz:2003vj} where the leading growth is $\mathcal{O}(s^3)$ and the cutoff is raised to $\Lambda_3$.\footnote{Recent work \cite{Bonifacio:2018vzv,Bonifacio:2018aon,Bonifacio:2019mgk} has demonstrated that $\Lambda_3$ is the maximum cutoff scale even in the presence of an arbitrary number of lower-spin particles.} Note, also, that the divergent high energy behavior depends on the particle mass, signifying an IR dependence of the UV cutoff. These properties have been verified by explicit computation \cite{Aubert:2003je,Cheung:2016yqr}.

In contrast, for theories where massive spin-2 particles arise from a compactified extra dimension, the scattering amplitudes must grow far less rapidly with energy. In such theories, the massless five-dimensional (5D) graviton field is decomposed into a sum of harmonic functions of the compactified internal space weighted by 4D spin-2 KK fields \cite{Kaluza:1921tu,Klein:1926tv,Appelquist:1987nr}.  The UV behavior of the properly normalized dimensionless 5D graviton scattering amplitude  in the underlying theory behaves like $s^{3/2}/M^3_{5}$, where $M_5$ is the 5D Planck scale.\footnote{The Feynman amplitude for $2\to 2$ scattering in 5D has units of (mass)$^{-1}$ and, compared to 4D, an additional factor of energy arises in the 5D partial wave expansion \cite{Soldate:1986mk,Chaichian:1987zt}.} Because the high energy behavior of the 4D scattering amplitudes must be consistent with the 5D theory, terms in the scattering amplitude that grow as $s^5$ (or even as $s^3$) must cancel among themselves. This cancellation is difficult to demonstrate in practice because of the complicated interaction vertices arising from the EH action.\footnote{The cancellations also make it impossible to use power-counting to analyze the continuum interacting KK theory as done in Refs.  \cite{Schwartz:2003vj,Randall:2005me,Gallicchio:2005mh}: the full theory has cancellations between different individual contributions, and a complete scattering amplitude calculation (as presented here) is needed to understand the high-energy behavior.}

Here we demonstrate explicitly how the needed cancellations occur both in the case of a torus where the internal space is flat and in the case of RS1 \cite{Randall:1999ee} (a slice of AdS$_5$) where the internal space has constant negative curvature; in the latter case,  compactification provides an additional dimensionful scale  \cite{ArkaniHamed:2000ds}.

\section{Orbifolded Torus}

Consider the 5D orbifolded torus (5DOT). The relevant 5D EH action is
\begin{equation}
S= \frac{2}{\kappa^{2}}\int d^{4} x \,d y  \sqrt{\operatorname{det} G_{MN}}\, R
\end{equation}
where $x^\mu$ are the coordinates of the four non-compact dimensions; $y\in [-\pi r_c,+\pi r_c]$ is the coordinate of the compact internal space; $G_{MN}$ and $R$ are the five-dimensional metric and Ricci scalar respectively; and the dimensionful coupling $\kappa=2/M^{-3/2}_5$ is the weak field expansion parameter fixed by the 5D Planck scale $M_5$. The KK theory relates the 4D and 5D Planck scales according to $M_{Pl}^{2} = 2\pi r_c M_{5}^{3} $.

Imposing an orbifold symmetry, the 5D metric then equals
\begin{equation}
G_{M N}=\left( \begin{array}{cc}{e^{-\kappa \hat{r}} / \sqrt{6}\left(\eta_{\mu \nu}+\kappa \hat{h}_{\mu \nu}\right)} & {0} \\ {0} & {-(1+\hat{r} / \sqrt{6})^{2}}\end{array}\right)~.
\end{equation}
where the 5D graviton field $\hat{h}(x,y)$ and 5D radion field $\hat{r}(x,y)$ are even functions under the orbifold reflection $y\rightarrow -y$. The tensor $\eta_{\mu\nu}$ is the usual  $4\times 4$ `mostly-minus' Lorentz metric $\text{diag}(+1,-\vec{1})$. This particular $G_{MN}$ parameterization renders all kinetic and mass terms automatically canonical. To calculate the scattering amplitudes, we obtain the terms describing 5D 3-point and 4-point couplings by expanding the EH Lagrangian to order $\kappa^2$. We perform this algebraically-intensive expansion using a new diagrammatic technique; this technique and the subsequent integration-by-parts reduction are automated in a way we will detail in a future publication.

KK decomposition replaces a 5D field $\hat{f}(x,y)$ with a complete sum of internal space harmonic wavefunctions $\psi_n(y)$ weighted by 4D fields $\hat{f}^{(n)}(x)$. Because the present internal space is flat and orbifolded, the wavefunctions are cosines (\`{a} la traditional Fourier decomposition) and each 4D `KK mode' $\hat{f}^{(n)}(x)$ may be labeled by a `KK number' $n$ equal to how many nodes its associated wavefunction has across $y\in[0,+\pi r_c]$. Following this procedure, the 5D graviton field $\hat{h}$ yields infinitely-many massive spin-2 KK modes with masses $m_n = n/r_c$ ($n > 0$) and one massless spin-2 KK mode which is identified with the 4D graviton ($n = 0$). Decomposing the radion is more straightforward: in a suitable gauge, the 5D radion field\footnote{The radion's VEV determines the size of the internal space. Any realistic theory must include a mechanism to stabilize (see, for example, \cite{Goldberger:1999uk}) this size, and in doing so give mass to the radion, which we consider in a subsequent work.} is constant across the internal space and satisfies $\hat{r}(x,y)\equiv \hat{r}(x)$ \cite{Callin:2004zm}. Consequently, its KK decomposition contains only a single massless spin-0 KK mode ($n=0$), the radion. From here on KK mode will refer to a massive spin-2 state, i.e. a mode with nonzero KK number.

\begin{figure}
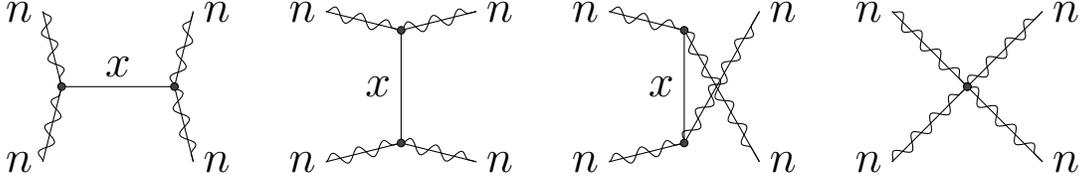

 $\sD{0.5}\hspace{15 pt}\tD{0.5}\hspace{15 pt}\uD{0.5}\hspace{15 pt}\sgD{0.5}$
\caption{Feynman diagrams contributing to $nn \to nn$ level spin-2 KK boson scattering. In the orbifold torus model, the intermediate states $x$ include the radion, the massless graviton, and the KK-mode at level $2n$.}
\label{fig:fd}
\end{figure}

By integrating the 5D EH Lagrangian over the internal space, we construct an effective 4D Lagrangian $\mathcal{L}^{(\text{eff})}_{4D} \equiv \int dy\hspace{5 pt}\mathcal{L}_{5D}$. The previously-attained 3-point and 4-point interactions between 5D fields become 3-point and 4-point interactions between various KK modes proportional to integrals of products of wavefunctions. For this flat internal space, discrete KK momentum conservation restricts the non-zero interaction vertices, {\it e.g.} a 3-point vertex attached to modes with KK numbers $l,m,n$ is only nonzero when $l = |m\pm n|$.

As an explicit example, consider the tree-level elastic scattering amplitude $\mathcal{M}$ of KK modes $(n,n)\rightarrow(n,n)$ and its expansion for large $s$. Due to KK momentum conservation, this amplitude has contributions arising only from the exchange of the KK mode at level $2n$, and the massless graviton and radion states (which yield   $t$- and $u$-channel IR divergences) as shown in Fig. \ref{fig:fd}. The first three combinations we consider are labeled by the relevant exchange particle, i.e. whether it is the $2n^{\rm th}$ KK mode, the graviton, or the radion; these sums of $s$-, $t$-, and $u$-channel exchange diagrams are labeled $\mathcal{M}_{2n}$,  $\mathcal{M}_{0}$, and $\mathcal{M}_{radion}$, respectively. The fourth combination consists solely of the 4-point contact interaction diagram $\mathcal{M}_{contact}$.  Up to second order in coupling $\kappa$, these diagrams form a diffeomorphism-invariant set. We calculate
  \begin{footnotesize}
\begin{eqnarray}
\mathcal{M}&=&\mathcal{M}_{2n}+\mathcal{M}_{0}+\mathcal{M}_{radion} + \mathcal{M}_{contact} \nonumber  \\
&\equiv&\sum_{k=-\infty}^{+5} \overline{\mathcal{M}}^{(k)} \cdot s^{k}~. \label{eq:Laurent}
\end{eqnarray}
\end{footnotesize}
%
and present the results for each class of diagrams in Table \ref{tab:nn2nn}.
By including all intermediate states we find (here $\theta$ is the center-of-mass scattering angle)
\begin{footnotesize}
\begin{equation}
\begin{aligned} 
\overline{\mathcal{M}}^{(5)} &=\overline{\mathcal{M}}^{(4)}=\overline{\mathcal{M}}^{(3)}=\overline{\mathcal{M}}^{(2)}=0 \\ \overline{\mathcal{M}}^{(1)}(\theta) &=\frac{3 \kappa^{2}}{256 \pi r_{c}}[7+\cos (2 \theta)] \csc ^{2} \theta ~
.\end{aligned}
\label{eq:torus-results}
\end{equation}
\end{footnotesize}
As anticipated, the amplitude does not grow like $s^5$ (or even $s^3$) despite individual contributions growing as fast as $s^5$. Instead, there are cancellations\footnote{Note that the radion contributes at ${\cal O}(s^3)$ as shown in \cite{Schwartz:2003vj}. However, if the theory is truncated below level $2n$, the $2n$th KK mode is absent and its contributions from the second row of Table \ref{tab:nn2nn} are not included.  Thus, the total amplitude in the truncated theory grows like ${\cal O}(s^5)$  -- not like ${\cal O}(s^3)$ as \cite{Schwartz:2003vj} had suggested.}  which lead to the total amplitude's growing only like $s$. Note the amplitude is proportional to $\kappa^2 /\pi r_c=8/M^2_{Pl}$, and is hence suppressed by the 4D Planck scale.

Additional calculations confirm cancellations that tamp growth down to $\mathcal{O}(s)$ for other $2 \to 2$ scattering processes as well, including processes like $(1,4) \to (2,3)$ to which the radion and graviton cannot contribute due to KK number conservation.  For processes lacking $t$- and $u$-channel IR divergences, we can directly compute the properly normalized partial-wave helicity  amplitude  \cite{Jacob:1959at}
\begin{footnotesize}	
\begin{equation}
a_{\lambda_{a} \lambda_{b} \rightarrow \lambda_{c} \lambda_{d}}^{J}=\frac{1}{32 \pi^{2}} \int d \Omega \quad D_{\lambda_{i} \lambda_{f}}^{J}(\theta, \phi) \mathcal{M}_{a \lambda_{b} \rightarrow \lambda_{c} \lambda_{d}}(s, \theta, \phi)~,	
\end{equation}
\end{footnotesize}
We find the largest (helicity-0, spin-0) partial wave amplitude has the leading behavior
\begin{footnotesize}
\begin{equation}
a^{J=0}_{00 \to 00} (14 \to 23) =\frac{s}{M_{\mathrm{Pl}}^{2}}  \ln \left(s r_{c}^{2}\right) + \ldots ~. \label{eq:unitarity}
\end{equation}
\end{footnotesize}
From this we conclude that 4D $2 \to 2$ scattering amplitudes from the 5DOT become large
at $s \simeq M^2_{Pl}$.

Finally, while each individual scattering amplitude grows only like $s$, as in the case of compactified Yang-Mills theory \cite{SekharChivukula:2001hz} there are coupled channels of the first $N$ KK modes whose scattering amplitudes grow like $Ns/M^2_{Pl}$. Following \cite{SekharChivukula:2001hz}, by identifying $N \propto \sqrt{s} r_c$ we recover the expected $s^{3/2}/M^3_{5}$ growth underlying five-dimensional gravity---and directly demonstrate the theory is valid up to a scale $\Lambda_{3/2}=M_{5}$ as suggested in \cite{Schwartz:2003vj}.

\begin{footnotesize}
\begin{table*}[t]
\begin{tabular}{|c|c|c|c|c|}
\hline
 & $ s^{5}$ & $s^{4}$ & $s^{3}$ & $s^{2}$  \\
 \hline
 ${\cal M}_{contact}$
& $ - \frac{ \kappa^{2} r_{c}^{7} [7 + c_{2\theta}] s_\theta^2 }{ 3072 n^{8} \pi } $
& $ \frac{ \kappa^{2} r_{c}^{5} [63 - 196\hspace{1 pt}c_{2\theta} + 5\hspace{1 pt}c_{4\theta}] }{ 9216 n^{6} \pi } $
& $ \frac{ \kappa^{2} r_{c}^{3} [-185 + 692\hspace{1 pt}c_{2\theta}  +5\hspace{1 pt}c_{4\theta}] }{ 4608 n^{4} \pi } $
& $ - \frac{ \kappa^{2} r_{c} [5 + 47\hspace{1 pt}c_{2\theta}] }{ 72 n^{2} \pi } $  \\
\hline
$ {\cal M}_{2n}$
& $ \frac{ \kappa^{2} r_{c}^{7} [7 + c_{2\theta}] s_\theta^{2} }{ 9216 n^{8} \pi } $
& $ \frac{ \kappa^{2} r_{c}^{5} [-13+ c_{2\theta}] s_\theta^{2} }{ 1152 n^{6} \pi } $
& $ \frac{ \kappa^{2} r_{c}^{3} [97 + 3\hspace{1 pt}c_{2\theta}] s_\theta^{2} }{ 1152 n^{4} \pi } $
& $ \frac{ \kappa^{2} r_{c} [-179 + 116\hspace{1 pt}c_{2\theta} - c_{4\theta}] }{ 1152 n^{2} \pi } $   \\
\hline 
$ {\cal M}_{0}$
& $ \frac{ \kappa^{2} r_{c}^{7} [7 + c_{2\theta}] s_\theta^{2} }{ 4608 n^{8} \pi } $
& $ \frac{ \kappa^{2} r_{c}^{5} [-9 + 140\hspace{1 pt}c_{2\theta} - 3\hspace{1 pt}c_{4\theta}] }{ 9216 n^{6} \pi } $
& $ \frac{ \kappa^{2} r_{c}^{3} [15 - 270\hspace{1 pt}c_{2\theta} - c_{4\theta}] }{ 2304 n^{4} \pi } $
& $ \frac{ \kappa^{2} r_{c} [175 +624\hspace{1 pt}c_{2\theta}  + c_{4\theta}] }{ 1152 n^{2} \pi } $   \\
\hline
${\cal M}_{radion}$
& $ 0 $
& $ 0 $
& $ - \frac{ \kappa^{2} r_{c}^{3} s_\theta^{2} }{ 64 n^{4} \pi } $
& $ \frac{ \kappa^{2} r_{c} [7 + c_{2\theta}] }{ 96 n^{2} \pi } $ \\
\hline
\hline
Sum & $0$ & $0$ & $0$ &  $0$ \\
\hline 
\end{tabular} 
\caption{Cancellations in the $(n,n) \to (n,n)$ 5DOT amplitude, where $\theta$ is the center-of-mass scattering angle and $(c_{n\theta},s_{n\theta}) = (\cos n\theta,\sin n\theta)$.  \label{tab:nn2nn}}
\end{table*}
\end{footnotesize}

\section{Anti-deSitter Space}

Next consider the analogous calculation in RS1 \cite{Randall:1999ee}. RS1 is a  truncated  and orbifolded Anti-de-Sitter space (AdS$_5$), bounded on either end by UV (Planck) and IR (TeV) branes.  Bulk and brane cosmological constants are added to the action to ensure the effective 4D background remains flat.\footnote{Here we address 5D gravity and ignore matter.}
The following RS1 metric generalizes the earlier 5DOT metric (which is recovered by taking $kr_c\rightarrow 0$ with finite $r_c$) \cite{Rattazzi:2003ea}
\begin{footnotesize}
\begin{eqnarray}
G_{M N} &=&\left( \begin{array}{cc}{e^{-2(k|y|+\hat{u})}\left(\eta_{\mu \nu}+\kappa \hat{h}_{\mu \nu}\right)} & {0} \\ {0} & {-(1+2 \hat{u})^{2}}\end{array}\right) \quad  \nonumber \\ 
\hat{u} &\equiv &\frac{\kappa \hat{r}}{2 \sqrt{6}} e^{+k\left(2|y|-\pi r_{c}\right)}~.
\end{eqnarray}
\end{footnotesize}
and is similarly canonical by construction. The new parameter $k$ has dimensions of mass and determines the curvature of the internal AdS$_5$ space.


In the `large $kr_c$ limit' ($kr_c\gtrsim 5$), the KK mode masses equal $m_{n}=k x_{n}e^{-kr_{c}\pi}$, where $x_{n}$ are zeroes of the Bessel function of the first kind. The location of the IR (TeV) brane determines an emergent scale $\Lambda_\pi \equiv M_{Pl}e^{-kr_{c}\pi}$ that controls the radion and KK mode coupling strengths. $\Lambda_\pi$ is exponentially suppressed relative to the 4D Planck scale  that determines graviton couplings ($M_{Pl}^2 = M_5^3/k$ at large $kr_c$). As we will show directly massive spin-2 scattering amplitudes in RS1 are suppressed by $\Lambda_\pi$.


Computing massive spin-2 scattering amplitudes in RS1 proceeds much like in the 5DOT, but with fewer conveniences ({\it e.g.}, see \cite{Davoudiasl:2001uj}). Since the internal space is curved, the harmonic functions are related to Bessel functions, but the resulting spectrum is similar to that of the 5DOT: a massless radion and graviton, and a tower of massive spin-2 KK states labeled by the number of nodes across the internal space. However, in RS1 there is no analog of KK momentum conservation, and so there are nonzero 3- and 4-point interactions between almost all combinations of 4D particles. Furthermore, the overlap integrals that accompany these interactions (containing three or four wavefunctions each) cannot be performed analytically. Investigating an RS1 scattering amplitude therefore requires accurate evaluation of the relevant highly-oscillatory wavefunctions and their overlap integrals. This difficulty is amplified by the large number of terms in each contribution: every intermediate KK mode contributes over 9300 terms to the scattering amplitude even before we substitute polarizations and momenta or expand in powers of energy -- then we must sum over all intermediate KK modes.

Consider the KK scattering amplitude $(1,1) \to (1,1)$ in RS1, and its expansion in energy per Eqn. (\ref{eq:Laurent}). Because KK momentum is not conserved in RS1, all KK modes contribute as intermediate states to this amplitude. In practice, therefore, we study the convergence of the amplitude as a function of $N_{max}$, the maximum KK level included as an intermediate state. From this perspective, we verify that cancellations in RS1 proceed just as they do in the 5DOT. In particular, we find that the contribution of the $N$th intermediate KK mode to $s^k$-like growth of the scattering amplitude scales like $1/N^{2k+2}$ for $k\in\{2,3,4,5\}$. By truncating at level $N_{max} \gae 10$ and summing over the states of higher mode number, we find the residual amplitude therefore scales like
\begin{equation}
\overline{\mathcal M}^{(k)}_{N_{max}} \propto {\cal O}\left( \frac{1}{N_{max}^{2k+1}}\right)~,
\end{equation}
for each $k\in\{2,3,4,5\}$ and these contributions all vanish in the $N_{max} \to \infty$ limit. By contrast, ${\mathcal M}^{(1)}$ converges to a finite result and the leading contribution to the amplitude scales like $s^1$ as expected.

We also find that the angular dependence of $\overline{\mathcal M}^{(1)}(\theta)$ is exactly the same as in the toroidal case. Dividing $\overline{\mathcal{M}}^{(1)}$ by its toroidal equivalent (with fixed $M_{Pl}$ and $m_1$), we can then scale from Eqn. (\ref{eq:unitarity}) to estimate the scale of validity of this 4D RS1 EFT calculation. We have done so for a number of different scattering amplitudes, and in all cases we find the 4D scattering amplitudes become strong at an energy scale $\sqrt{s} \simeq \Lambda_\pi$ -- verifying directly that the cutoff scale for the RS1 effective field theory, as determined by the exclusive scattering amplitudes, is controlled by $\Lambda_\pi$.

\section{Discussion}

We have reported on the first complete calculations of the tree-level scattering amplitudes of the longitudinal modes of massive spin-2 Kaluza-Klein states. Since completing this work, we have found an alternative way to demonstrate the cancellation both for the flat and curved internal spaces, via sum rule techniques \cite{Chivukula:2019zkt}; and other groups have likewise since put forth sum rule approaches for Ricci flat internal spaces \cite{Bonifacio:2019ioc}. Details of the computations presented will be given in forthcoming work, which will also address related issues such as the effects of radion stabilization, the inclusion of matter fields, and phenomenological impacts.  

This material is based upon work supported by the National Science Foundation under Grant No. PHY-1915147 .



%

\end{document}